\documentclass[12pt,preprint]{aastex}

\usepackage{color}           
\usepackage{ulem}            

\slugcomment{}

\shorttitle{Electron Acceleration around the X-point}
\shortauthors{Narukage et al.}

\begin{document}


\title{Evidence of Electron Acceleration around the Reconnection X-point in a Solar Flare}


\author{Noriyuki Narukage and Masumi Shimojo\altaffilmark{1}}
\affil{National Astronomical Observatory of Japan, Mitaka, Tokyo 181-8588, Japan}
\email{noriyuki.narukage@nao.ac.jp}

\and

\author{Taro Sakao\altaffilmark{2}}
\affil{Institute of Space and Astronautical Science, Japan Aerospace Exploration Agency, Sagamihara, Kanagawa 252-5210, Japan}


\altaffiltext{1}{Department of Astronomical Science, School of Physical Sciences,
                 The Graduate University for Advanced Studies, Mitaka, Tokyo 181-8588, Japan}
\altaffiltext{2}{Department of Space and Astronautical Science, School of Physical Sciences,
                 The Graduate University for Advanced Studies, Sagamihara, Kanagawa 252-5210, Japan}


\begin{abstract}
Particle acceleration is one of the most significant features that are ubiquitous
among space and cosmic plasmas. It is most prominent during flares in the case of the Sun,
with which huge amount of electromagnetic radiation and high-energy particles are expelled
into the interplanetary space through acceleration of plasma particles in the corona.
Though it has been well understood that energies of flares are supplied by the mechanism
called magnetic reconnection based on the observations in X-rays and EUV with space telescopes,
where and how in the flaring magnetic field plasmas are accelerated has remained unknown
due to the low plasma density in the flaring corona.
We here report the first observational identification
of the energetic non-thermal electrons
around the point of the ongoing magnetic reconnection (X-point);
with the location of the X-point identified by soft X-ray imagery and
the localized presence of non-thermal electrons identified from imaging-spectroscopic data at two microwave frequencies.
Considering the existence of the reconnection outflows that carries both plasma particles and magnetic fields out from the X-point,
our identified non-thermal microwave emissions around the X-point indicate that
the electrons are accelerated around the reconnection X-point.
Additionally, the plasma around the X-point was also thermally heated up to 10~MK.
The estimated reconnection rate of this event is $\sim$~0.017.
\end{abstract}


\section{Introduction}
\label{sec:Introduction}

Particle acceleration is ubiquitously observed among space and cosmic plasmas,
including those around the Earth's magneto-tail \citep[{\it e.g.},][]{oie02}, in supernova remnants \citep[{\it e.g.},][]{koy95},
and even in distant galaxies \citep[{\it e.g.},][]{mus77}.
For our Sun, acceleration of plasma particles is most prominent in flares,
with which vast amounts of electromagnetic radiation from accelerated particles \citep[{\it e.g.},][]{sak96, yok02},
as well as the particles themselves \citep[{\it e.g.},][]{lin96}, are expelled into interplanetary space.
Observations from space in the past few decades have established that it is magnetic reconnection
during flares that converts magnetic energy in the corona into the kinetic and thermal energies
of the plasma particles \citep[{\it e.g.},][]{tsu92, mas94, yok01, har11, ima13, su13}.
It is thus expected that particle acceleration in flares
should also be closely related to the magnetic reconnection process through
which as much as half of the liberated energy is converted into acceleration of particles \citep{lin76}.

There have been a number of theoretical studies made so far that have considered various portions of the reconnecting magnetic structure
as the site of electron acceleration in solar flares.
These include, inside the closed magnetic loop formed by reconnection \citep[{\it e.g.},][]{fle08},
in the magneto-hydrodynamic fast-shock structure expected to be formed above the closed loop \citep[{\it e.g.},][]{tsu98},
in the magnetic cusp region where field lines are contracting downward \citep[{\it e.g.},][]{som97},
and around the X-point (or in the current sheet) \citep[{\it e.g.},][]{lit96,pri06,dra06,oka10}.
Meanwhile, some observations have addressed possible site of the electron acceleration.
\citet{asc96} studied timing relationship among hard X-ray emissions in different energies 
and quantitatively estimated locations of electron acceleration region as above the flaring loops.
\citet{sak98} argued that electron acceleration should take place at the reconnection site
which is a common place among the two different magnetic field configurations 
identified from foot point motions of hard X-ray sources during the impulsive phases.
These observations gave us the supposition that the particle acceleration would occur above the flaring loops.

However, there have not been any observations that pinpoint where and how in the flaring magnetic field plasmas are accelerated.
This is partly due to the fact that magnetic reconnection in flares takes place
in the high corona where plasma density is low, and
hard X-ray fluxes from energetic non-thermal electrons, which are emitted by Bremsstrahlung
(interaction with ambient coronal plasma), are too weak to be imaged by existing modulation-collimator type hard X-ray telescopes
whose dynamic range is not high, especially for the reconnection region.
Meanwhile, there is a possibility that microwaves emitted from the energetic non-thermal electrons
by gyro-syncrotoron mechanism (interaction with coronal magnetic fields) can be detected
even in the low density corona including the reconnection region.

In this paper, we report the first identification of
energetic non-thermal electrons around the reconnection X-point in a flare; with the location of the X-point
identified by soft X-ray imagery and the localized presence of non-thermal electrons identified
from imaging-spectroscopic data at two microwave frequencies.
This is a strong evidence of the electrons being accelerated around the X-point,
providing an observational clue toward understanding the mechanism of electron acceleration in solar flares.

\section{Observations}
\label{sec:Observations}

We present soft X-ray and microwave imaging observations, made with the Soft X-ray Telescope (SXT) \citep{tsu91}
aboard the {\it Yohkoh} satellite and the Nobeyama Radioheliograph \citep{nak94, tak97}, respectively,
of the 6 August 1999~UT flare that occurred from around 04:34~UT in NOAA active region 8647 on the west limb of the Sun
(S19W91; see Figure~\ref{fig:data}).
The X-ray flux detected by GOES satellite was almost flat at a level of $\sim$ C6.
The evolution of this flare was not seen in the GOES light curve,
probably because this flare was not large enough in soft X-ray flux
against the other eight active regions simultaneously located on the solar disk.
Hence, we track the evolution of this flare with the soft X-ray intensity
from the flaring loop derived from the {\it Yohkoh}/SXT data
as described in Section~\ref{sec:Observations in Soft X-Rays}.
Though we also check the hard X-ray data taken with the Hard X-ray Telescope (HXT) aboard the {\it Yohkoh} satellite,
the hard X-ray signal from this flare was very weak against the background level.
This is probably because the foot point of the flaring loop was located behind the solar limb.
Hence, we do not discuss the hard X-ray data in this paper.

\subsection{Location of the X-Point Identified with Soft X-Ray Imagery}
\label{sec:Observations in Soft X-Rays}

{\it Yohkoh}/SXT observed the evolution of the flare, from its very early stage, followed
by an ejection of a plasmoid structure as seen in Movie~1.
The corresponding large-scale feature was also observed in an extreme-ultraviolet (EUV) wavelength
(195 \AA; Movie~2) \citep{nar06}. From around 04:34~UT, the width of the portion connecting
the top of the bright soft X-ray loop and the plasmoid started to decrease with time,
which eventually led to the plasmoid ejection after $\sim$~04:50~UT (see Figure~\ref{fig:microwave}~(a)).
We note that the soft X-ray intensity from the bright soft X-ray loop derived from the SXT data continued to increase
as the width decreased while it turned into a decrease after the ejection, following the peak
in the soft X-ray flux (Figure~\ref{fig:microwave}~(b)). Such behavior in soft X-rays is consistent with
what is expected from the well-perceived reconnection picture for flares (Figure~\ref{fig:data}~(a)) \citep{shi11},
in which the observed decrease in width is attributed to reconnection inflow \citep{yok01, nar06}
towards the X-point while magnetic reconnection is in progress.
From the soft X-ray image at the period of ongoing reconnection (Figure~\ref{fig:data}~(c)),
we have identified the X-like feature at the apex of the cusp shape structure,
{\it i.e.}, the smallest portion in width in the soft X-ray intensity contour map
(see the contours in Figure~\ref{fig:data} at 04:45:00~UT), as
the location of the reconnection X-point,
which is indicated by black circles in Figure~\ref{fig:data}~(b)--(d).
We consider that the X-point is located within the fixed black circles from the start of the reconnection to the plasmoid ejection,
because the location of the plasmoid gradually moved upward as the flare progressed, but
such moving distance is comparable to the diameter of the black circles until the plasmoid ejection at around 04:50~UT.
After the plasmoid ejection,
the X-point location can not be tracked, since the plasmoid was ejected outside of the field of view.
Additionally, the apex of the cusp shape structure shifted towards northeast direction after the plasmoid ejection.
Hence we move the black circle
along the yellow line in Figure~\ref{fig:data}~(b)--(d) to track the
location around the apex of the cusp shape structure
(around the edge of the current-sheet-like feature) as seen in Movie~3.
In the following, quantities referred to as those around the X-point correspond to those averaged over the black circle.

\subsection{Localized Presence of Non-Thermal Electrons Identified with Microwave Imaging-Spectroscopy}
\label{sec:Observations in Microwave}

\begin{figure}
\epsscale{.40}
\plotone{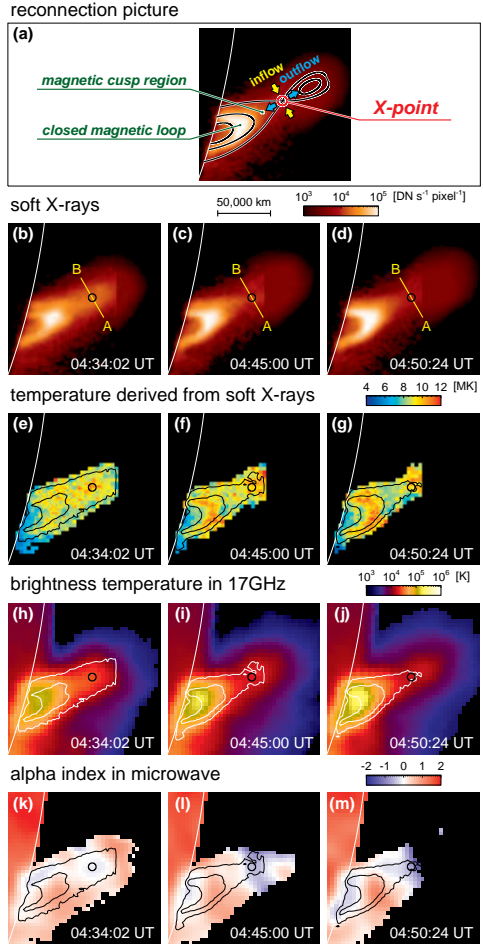}
\caption{Soft X-ray and microwave observations of the 6 August 1999 flare.
{\bf (a)} Schematic illustration of the reconnection picture for the flare.
{\bf (b)--(d)} Soft X-ray images taken with the AlMg filter of {\it Yohkoh}/SXT.
Each of them is a composite frame made from a pair of full-resolution (2.46$\arcsec$ pixel$^{-1}$) and
half-resolution (4.92$\arcsec$ pixel$^{-1}$) images.
The location of the reconnection X-point is indicated by a black circle in each panel,
whose diameter is set as 10$\arcsec$ ($\sim$~7,300~km) which is approximately the beam size for the 17~GHz maps
shown in (h)--(j).
The yellow line is drawn to cross the center of the black circle (X-point) and
in perpendicular direction to the plasmoid ejection.
{\bf (e)--(g)} Electron temperature distribution derived with the filter ratio method from pairs of
soft X-ray images obtained with the AlMg and Be119 filters of {\it Yohkoh}/SXT.
Errors in the derived filter-ratio temperature are less than 15~\%.
{\bf (h)--(j)} 17~GHz microwave brightness temperature maps from Nobeyama Radioheliograph (NoRH).
{\bf (k)--(m)} Spatial distribution of power-law index of microwave flux density spectra (alpha index)
derived from NoRH 17~GHz and 34~GHz data.
Distribution of alpha index with its error less than 1.0 is displayed in each panel of (k)--(m).
Black or white contours in (e)--(m) trace soft X-ray intensity levels of 8,000 and 30,000~DN~s$^{-1}$ pixel$^{-1}$
for the full-resolution portion taken with the AlMg filter
(thus the vertical feature at the right end of the outer contour level in (e)--(m) is artificial).
The accuracy of the co-alignment between soft X-ray and microwave images is better than 2$\arcsec$.
In each panel, the white arc represents the solar limb. North is up and east is to the left.
\label{fig:data}}
\end{figure}

The flare was simultaneously observed in two microwave frequencies at 17~GHz and 34~GHz
with the Nobeyama Radioheliograph (NoRH). Figure~\ref{fig:data}~(h)--(j) give maps of the brightness temperature at 17~GHz
showing a similar spatial distribution as that in soft X-rays. In Figure~\ref{fig:data}~(k)--(m),
we present the spatial distribution of the power-law index of the microwave flux density spectra
(hereafter referred to as ``alpha index" \citep{dul85}) derived from the brightness temperature maps at 17~GHz and 34~GHz.
As the 34~GHz maps suffered from aliasing patterns from image synthesis,
we carefully removed their effect before evaluating the alpha index;
the procedure for which is detailed in Appendix~\ref{sec:Calibration of 34 GHz data}.

The meaning of alpha index value is summarized in Table~\ref{tbl:alpha index}, based on Figure~2 in \citet{dul85}.
The sign of alpha index changes from positive to negative or zero at the frequency where $\tau \sim 1$ ($\tau$ is the optical depth).
The degree of alpha index is determined by the emission mechanism of the microwave.
As derived by \citet{dul85}, for the optically thin gyro-synchrotron emission, the alpha index $\alpha$ is
related to the spectral index of the accelerated electrons $\delta_{\mu}$ in such a way that
$\delta_{\mu} = (1.22 - \alpha) / 0.9$.
Note that there is a difference between the electron spectral indices
derived from microwaves $\delta_{\mu}$ and hard X-rays $\delta_{X}$,
since each wavelength is emitted from the electrons in different energy ranges
by different emission mechanism ({\it i.e.}, gyro-syncrotoron and Bremsstrahlung, respectively).
According to \citet{asa13}, the difference between
$\delta_{X}$ derived from {\it Yohkoh}/HXT
and $\delta_{\mu}$ from NoRH
is $\delta_{X} - \delta_{\mu} \sim 1.6$ for 8 flare examples.

We note that in Figure~\ref{fig:data}~(k)--(m), and also in Movie~4, there is a localized distribution of
negative alpha index around the X-point for the period of ongoing reconnection
(at 04:45~UT; Figure~\ref{fig:data}~(l)) and around the timing of the plasmoid ejection (at 04:50~UT; Figure~\ref{fig:data}~(m)),
while the alpha index is around zero in the early stage of the flare (at 04:34~UT; Figure~\ref{fig:data}~(k)).
The degree of circular polarization at 17~GHz for the region above the bright soft X-ray
(as well as microwave) loop was at most $\sim$~2~\% so the possibility of gyro-resonance emission
for this region can be safely ruled out. With this possibility excluded,
the negative values of alpha index around the X-point immediately indicate, by themselves,
that the microwaves there originate from optically-thin (gyro-)synchrotron emission by non-thermal,
{\it i.e.}, accelerated, electrons of (mildly-)relativistic energies
(see Table~\ref{tbl:alpha index}).
We note that, in reality, thermal electrons should also coexist in the corona,
and the observed alpha index contains contribution from both thermal and non-thermal electrons along the line-of-sight.
Even in this case, the negative alpha index indicates the existence of non-thermal electrons,
because only the optically-thin non-thermal gyro-syncrotoron emission can cause the negative alpha index
except for the optically-thin gyro-resonance emission that was excluded as mentioned above (see Table~\ref{tbl:alpha index}),
although $\delta_{\mu}$ cannot be estimated correctly from this alpha index.
The amount of the non-thermal electrons is discussed in Section~\ref{sec:Non-Thermal Component}.
The alpha index of $\sim$~0 in the early stage of the flare is likely due to optically-thin
Bremsstrahlung emission from thermal electrons at the initial stage of the acceleration process.
Figure~\ref{fig:microwave}~(c) gives the temporal evolution of the alpha index around the X-point.
As the reconnection proceeds towards the plasmoid ejection, the alpha index decreases
to larger negative values, increasing significance of the presence of non-thermal emission there.
After the ejection, the alpha index in the black circles of Figure~\ref{fig:data}
(that show the location of the X-point before the plasmoid ejection, and now around the edge of
the current-sheet-like feature) returned towards zero, suggesting thermalization there or
evacuation of the accelerated electrons from the region.

\section{Estimate of the Non-Thermal Component}
\label{sec:Non-Thermal Component}

In addition to the non-thermal (gyro-)synchrotron component discussed above,
the observed microwave brightness temperature contains another component from Bremsstrahlung emission
by thermal electrons along the line-of-sight. Since we can infer the temperature and column emission measure
of the thermal electrons from the SXT images with the filter ratio method \citep{tsu91, tak11} as shown in Figure~\ref{fig:sxr},
the brightness temperature at 17~GHz expected from soft X-ray-emitting thermal plasmas,
$T_\mathrm{B}^\mathrm{Th}$, can then be calculated \citep{dul85} as shown in Figure~\ref{fig:microwave}~(d).
By subtracting $T_\mathrm{B}^\mathrm{Th}$ from the measured brightness temperature
at 17~GHz, we obtain the residual brightness temperature $T_\mathrm{B}^\mathrm{NT}$ that may be a good measure of the emission
from non-thermal electrons. The ``non-thermal brightness temperature" $T_\mathrm{B}^\mathrm{NT}$ thus derived
around the X-point rapidly increased to $\sim$~7,000~K in the early stage of the reconnection process
at around 04:34~UT (see Figure~\ref{fig:microwave}~(e)). It stayed around this level throughout the reconnection process
until the ejection of the plasmoid. This may indicate that the electron acceleration process continued
to be in operation during this period. Since the temporal evolution of the alpha index and $T_\mathrm{B}^\mathrm{NT}$
clearly synchronize with the reconnection process, it is most likely that the acceleration process
of electrons around the X-point closely traces the progress of magnetic reconnection.
After the ejection, we see drastic decrease in $T_\mathrm{B}^\mathrm{NT}$. This indicates that the electron acceleration either ceased,
or weakened significantly around the edge of the current-sheet-like feature
where the X-point was located before the plasmoid ejection.

\section{Discussion}
\label{sec:Discussion}

We have shown that there is microwave emission from non-thermal electrons most clearly seen
around the X-point. This indicates that there are energetic electrons present
around the X-point during the course of the magnetic reconnection.
Meanwhile, the soft X-ray observations of the flare were in good agreement, in a morphological sense,
with the canonical reconnection picture (Figure~\ref{fig:data}~(a)) from which we expect,
in addition to the inflow, bi-directional outflow of plasma particles and magnetic fields
away from the X-point towards the bright soft X-ray loop and in the opposite direction \citep{shi11}.
Assuming the presence of the expected reconnection outflow, it would be reasonable
to conclude that the energetic non-thermal electrons are supplied from, namely, accelerated at,
the region around the X-point rather than assuming that they travel
from a region lower in altitude than the X-point against the counter-streaming downward outflow,
or that they come from a region higher in altitude beyond the X-point
against the counter-streaming upward outflow from the X-point.
The observed decreasing trend in the column emission measure (Figure~\ref{fig:sxr}~(c)),
{\it i.e.}, a decrease in density of soft-X-ray-emitting thermal electrons,
would be due to the ambient electrons being swept away from the X-point
by the reconnection outflow and/or that they departed from thermal equilibrium
due to collisions with accelerated electrons \citep{kru10}.
Furthermore, as seen in Movie~4, the negative alpha signal traveled from around the X-point
to the foot point of the flare loop along the northern outer edge of the flare loop
around the timing of the plasmoid ejection. This gives further support that the accelerated electrons
manifesting themselves as the negative alpha signal are originated from around the X-point,
although the traveling negative alpha signal does not immediately correspond to the individual energetic electrons.
Hence, we argue that electron acceleration around the X-point is the initial (first-stage) acceleration
in a flare that immediately follows the onset of reconnection.

We further take a closer look into the region around the X-point.
The spatial extent of the negative alpha index in the lateral direction
(parallel to the solar surface) is possibly caused by the inclined configuration
of the flaring loop to the line-of-sight as reported by \citet{nar06}.
Meanwhile, the length of the localized negative alpha index,
{\it i.e.}, the region where the energetic non-thermal electrons exists,
has a spatial extent of $\sim$~20,000 km along the direction of the height, and is located
at the top-most of the cusp structure,
for the period of ongoing reconnection (see Figure~\ref{fig:data}~(l)).
Considering this spatial extent along the direction of the height,
the possible scenarios of the electron acceleration around the X-point are follows:
(i) An X-point alone can produce energetic electrons
with the inductive reconnection electric field \citep[{\it e.g.},][]{pri06}.
Such energetic electrons can easily escape from the X-point, and emit microwaves from a spatially extended region as seen in our data.
(ii) Multi-island coalescence along a single current sheet should also be considered as an electron acceleration scenario \citep[{\it e.g.},][]{oka10}.
(iii) Fragmented islands scenario with multiple current sheets \citep[{\it e.g.},][]{dra06, shi01} is also possible,
since, considering the spatial resolution of our microwave data ($\sim$ 10$\arcsec$),
the multiple current sheets cannot be resolved, and the non-thermal signals in microwaves should be observed
as only one source as seen in the presented data.

We note that the hot thermal electrons ($>$ 10~MK) spatially coexists with the energetic non-thermal electrons
(Figure~\ref{fig:data}~(e)--(g)) around the X-point.
Since the coronal temperature below the X-point are cooler than around the X-point,
we can say that the hot plasmas around the X-point is not heated by the thermal conduction from the plasmas located below the X-point.
These high temperature plasmas are directly created by the released magnetic energy
that would also accelerate the electrons, and/or are the result of the thermalization of the accelerated electrons.

In our flare, we also find microwave non-thermal sources other than around the X-point
(see the negative alpha index distribution in Figure~\ref{fig:data}, especially in Figure~\ref{fig:data}~(m)).
One is located at the foot points of the flaring loop.
This negative alpha index is also due to the non-thermal emission, not by the gyro-resonance emission,
since the sunspots were occulted by the solar disk in this flare.
The other non-thermal source is distributed above the top of the flaring loop extending from the X-point region.
These two kinds of observed microwave non-thermal sources became noticeable
around the timing of the plasmoid ejection (see Figure~\ref{fig:data}~(m)).
This is consistent with the timing where intensity enhancement in the hard X-rays
was observed around the timing of plasmoid ejections \citep[{\it e.g.},][]{ohy98}.
Hence, these two kinds of the sources may have close relationship with
the well known hard X-ray non-thermal sources, {\it e.g.}, double-foot-point sources \citep[{\it e.g.},][]{sak96}
and above-the-loop-top source \citep[{\it e.g.},][]{mas94}.
Considering the spatially-smooth distribution of the negative alpha index extending from the X-point region
(larger negative alpha index $\sim -1$) to above-the-loop-top region (smaller negative alpha index $\sim 0$)
at the timing when the microwave non-thermal sources became noticeable
(see Figure~\ref{fig:data}~(m)), it is reasonable to expect that
no major acceleration site other than around the X-point exists,
although some additional sub-accelerations might occur in other locations.
The hard X-ray above-the-loop-top source \citep[{\it e.g.},][]{mas94}
may appear as a lower part of the spatial distribution of non-thermal electrons from the X-point region to above the loop top
manifesting themselves in microwaves (with negative alpha index),
and is detected due to the denser coronal plasmas near the flaring loop
with which Bremsstrahlung hard X-rays are efficiently emitted.

Meanwhile, the reconnection rate can be estimated to be $\sim$~0.017 from the data set
of following parameters considering the line-of-sight effect \citep{nar06}: the inflow velocity ($\sim$~16.3 km s$^{-1}$)
derived from Figure~\ref{fig:microwave}~(a), the foot point expanding velocity ($\sim$~3.3 km s$^{-1}$)
of the flare loop from X-ray data, the averaged magnetic field strength at the sunspot ($\sim$~98 Gauss)
from the magnetogram, and the coronal column emission measure ($\sim 10^{28.4}$ cm$^{-5}$
that corresponds to the number density of $10^{9.3}$ cm$^{-3}$ for the line-of-sight depth of 60,000 km)
from the soft X-ray data with the filter ratio method. This reconnection rate supports the Petschek type reconnection
against Sweet-Parker type \citep{pri82}.

The solar corona provides a unique opportunity for investigating the acceleration process associated with
magnetic reconnection, not only in that the entire view of the reconnecting magnetic structure can be obtained by imagery,
but also because the spatio-temporal evolution of any sources with certain spectral features
in the dynamically-evolving reconnecting structure can be traced by imaging-spectroscopic approach.
With the present result being the first step, progress in theoretical investigations as well as future observations
such as from Atacama Large Millimeter/submillimeter Array (ALMA) \citep{bas02} with very high spatial resolution,
mirror-focusing hard X-ray \citep[FOXSI;][]{kru13}
and soft X-ray \citep{sak12} telescopes with photon-counting capabilities
should help us solve the long-standing questions about particle acceleration in solar flares.

\acknowledgments

This work made extensive use of the {\it Yohkoh} Legacy Data Archive at Montana State University, U.S.,
which is supported by NASA.
The authors thank D. H. Brooks for polishing the text.
The {\it Yohkoh} mission was developed, launched and operated by ISAS/JAXA, Japan,
with NASA and SERC/PPARC (U.K.) as international partners.

\appendix

\section{Removal of the aliasing patterns in 34~GHz brightness temperature maps}
\label{sec:Calibration of 34 GHz data}

We here explain the method we devised for removing the aliasing patterns in synthesized images
at 34~GHz taken with the Nobeyama Radioheliograph (NoRH). NoRH is a radio interferometer dedicated
for solar observations at 17 GHz and 34 GHz microwave frequencies \citep{nak94, tak97}.
A set of raw signals from NoRH gives spatial Fourier components of the two-dimensional brightness distribution
across the full Sun, and it is necessary to reconstruct (synthesize) images from the set of raw data
to obtain a microwave map of the Sun. NoRH was designed to have its fundamental antenna spacing
(smallest antenna spacing) in such a way that it can measure Fourier components corresponding to
40~arcmin at 17~GHz. This implies NoRH is capable of synthesizing full-Sun microwave maps at 17~GHz free
from the aliasing effect while there are inevitable aliasing patterns present in synthesized 34~GHz maps
as illustrated in Figure~\ref{fig:34 G data}.
These aliasing patterns (pseudo brightness temperature distribution) drift as the azimuth and
elevation of the Sun seen from NoRH change with time (Movie~5).
In deriving the microwave alpha index from sets of 17~GHz and 34~GHz maps,
it was crucial to correctly remove the aliasing patterns in the 34~GHz maps.

In this study, we removed only ``aliasing pattern~1" (hereafter AP1) shown in Figure~\ref{fig:34 G data} and Movie~5,
since it is the only aliasing pattern component whose intense part passed over the flare site we analyzed (see Movie~5).
The effect of other aliasing patterns and residual random fluctuation in brightness temperature maps with image synthesis
will be discussed later.

For the removal of AP1, the motion of AP1 should be tracked. However, there was no intense feature
in AP1 suitable for tracking its motion. Hence, we first identified the motion of the pseudo flare site
produced by the aliasing effect in aliasing pattern~2 (indicated by a blue arrow in Figure~\ref{fig:34 G data} and Movie~5)
using correlation tracking of the real flare site (indicated by a red arrow in Figure~\ref{fig:34 G data} and Movie~5)
and the pseudo one. Then, by making use of the motion of the pseudo flare site and the fact that the motions of
aliasing patterns~1 and 2 are symmetric about the center of the solar disk (see Movie~5), we identified the motion of AP1.

In order to extract AP1 from the original 34 GHz map, we took the following approach.
As we know the relative motion between the real Sun and AP1, instead of looking into the drifting AP1
as we see in Movie~5, we fixed the aliasing pattern spatially and let the image of the real Sun drift
(Figure~\ref{fig:track and scan} and Movie~6).
At sufficient heights from the microwave disk of the Sun (shown as the green-colored zone in the left-hand side of Movie~6),
we can expect that any features seen above that height, {\it i.e.}, inside the green-colored zone, are of non-solar origin,
that is, originate from the aliasing. This way, AP1 can be extracted from the brightness distribution in the green-colored zone
by utilizing the scanning motion of the real solar disk with an assumption that AP1 was stable over the time of the observations.
The resultant patterns of AP1 are shown in green in the right-hand side of Figure~\ref{fig:track and scan} and Movie~6.
Note that each point of the aliasing pattern map gives a brightness temperature for AP1 at a combination of certain time and
location on the real Sun. For the X-point structure discussed in the Main Text, the location of AP1 that passes through
the X-point region (from 04:30:30~UT to 05:15:00~UT) is shown in red in Figure~\ref{fig:track and scan} and Movie~6.
An average of 60 consecutive brightness temperature maps (each with 10-seconds accumulation time, with a duration of 10 minutes)
was taken for determining the brightness at each spatial point in AP1. As can be seen in Movie~5,
AP1 drifts through the flare site before it moves away from the limb.
This means that AP1 determined in the green-colored zone lags behind the time when the corresponding portion of AP1
drifted through the flare site. In order to minimize the effect of any temporal variation in AP1,
when determining the brightness temperature at each spatial point in AP1, the first 60 maps after the point entered into
the green-colored zone were used. For the final phase of the flare (after 05:08:20~UT), less than 60 maps were available
because of a decrease in image quality of the synthesized 34~GHz maps after 05:38:30~UT (which were used to determine AP1
that passed through the flare site after 05:08:20~UT).
The relationship between the time that AP1 drifted through the flare site and that of determining AP1
in the green-colored zone is illustrated in Figure~\ref{fig:cal dataset}.

The brightness temperature distribution of AP1 thus derived was subtracted from the original 34~GHz map
for each epoch during the observations. Figure~\ref{fig:cal result} and Movie~7 illustrate pre- and post-removal of AP1
around the flare site at 34~GHz where we can see that the bright features in AP1 are successfully removed.
After the removal of AP1, however, there are still some flickering features caused by
(1) residual random fluctuation due to image synthesis,
(2) aliasing patterns other than AP1 (although weak in brightness) and
(3) any possible temporal variation in AP1.
Let us now evaluate these effects. Figure~\ref{fig:cal check} indicates the variation in brightness temperature of the aliasing patterns
along the red line in the right-hand panel of Figure~\ref{fig:track and scan}.
Note that the horizontal axis (with a label ``pixel") of this figure gives the location along the red line in Figure~\ref{fig:track and scan}
and also gives (with a label ``time") the time when that location of AP1 passed the site of the X-point.
For each point on the horizontal axis, the distribution in colors along the vertical axis gives the fractional distribution
of the brightness temperatures among the 60 maps used (until 05:08:10~UT; with fewer numbers of maps after then).
This fractional distribution reflects the stability of the brightness temperatures in the drifting aliasing pattern,
with the average and standard deviation for each point on the horizontal axis shown in the same figure.
The almost constant standard deviation of $\sim$~1,000~K across the entire horizontal axis location in the figure suggests that
random fluctuation due to image synthesis is the main cause of this standard deviation.
The amount of the standard deviation in the brightness temperature of AP1 was incorporated
when obtaining the real brightness temperature by subtracting AP1 from the original 34~GHz maps.
The error in the alpha index derived in Main Text (error bars in Figure~\ref{fig:microwave}~(c)) comes from this standard deviation.

So far we have used the pseudo brightness temperature from AP1, extracted from the 34~GHz maps in the time range of
04:57:00~UT--05:38:30~UT, for removing the effect of AP1 that passed the X-point between 04:30:30~UT and 05:15:00~UT.
There is typically a time gap of $\sim$~30 minutes present between the two
(the time gap is not constant due to the non-linear drifting motion of AP1 with respect to the real Sun).
In order to check whether the pseudo brightness temperature distribution of AP1 was stable enough over this time span,
another set of brightness distribution of AP1 was extracted, by utilizing the on-disk portion of the microwave Sun
(shown as a blue circle in the left-hand side of Figure~\ref{fig:track and scan})
where there was little activity (confirmed from EUV images; Movie~2).
In such a quiet region, an almost flat distribution of 34~GHz brightness temperatures at $\sim$~9,000~K is expected \citep{sel05}.
34~GHz maps from 04:05:00~UT to 04:50:40~UT were used for determining AP1 utilizing the on-disk quiet portion,
by subtracting the flat disk component at 9,000~K. This gives a pseudo brightness temperature distribution from AP1
that passed the X-point from 04:40:40~UT to 05:15:00~UT as illustrated in Figure~\ref{fig:cal dataset}.
The quality of the synthesized maps before 04:04:50~UT (which corresponds to the X-point passage before 04:40:30~UT) deteriorated,
hence they were excluded from the analysis. The resultant average pseudo brightness temperature distribution
(each derived from 60 consecutive maps closest in time to the X-point passage) is shown by a black line in Figure~\ref{fig:cal check}.
Throughout the period where AP1 of the pre-flare-site passage was available (after 04:47:30~UT),
the overlaid profile falls well within the range of the standard deviation, indicated by thin white lines in the figure.
Thus we conclude that the brightness temperature distribution of AP1 around the flare site can be considered stable during the flare,
within the range of standard deviation, supporting the validity of the method we used for removing AP1 from the 34~GHz maps of the flare.

\section{Captions about the supplementary movies}
\label{sec:movies}

{\bf Movie 1:}
Evolution of the flare observed in soft X-rays with the AlMg filter of {\it Yohkoh}/SXT.

{\bf Movie 2:}
Global evolution of the flare observed in EUV 195 {\AA}
with the Extreme ultraviolet Imaging Telescope (EIT) aboard {\it SOHO}.
North is up and east is to the left. The flare site is located in the south-western part of the Sun, on and above the solar limb.

\begin{figure}
\epsscale{.55}
\plotone{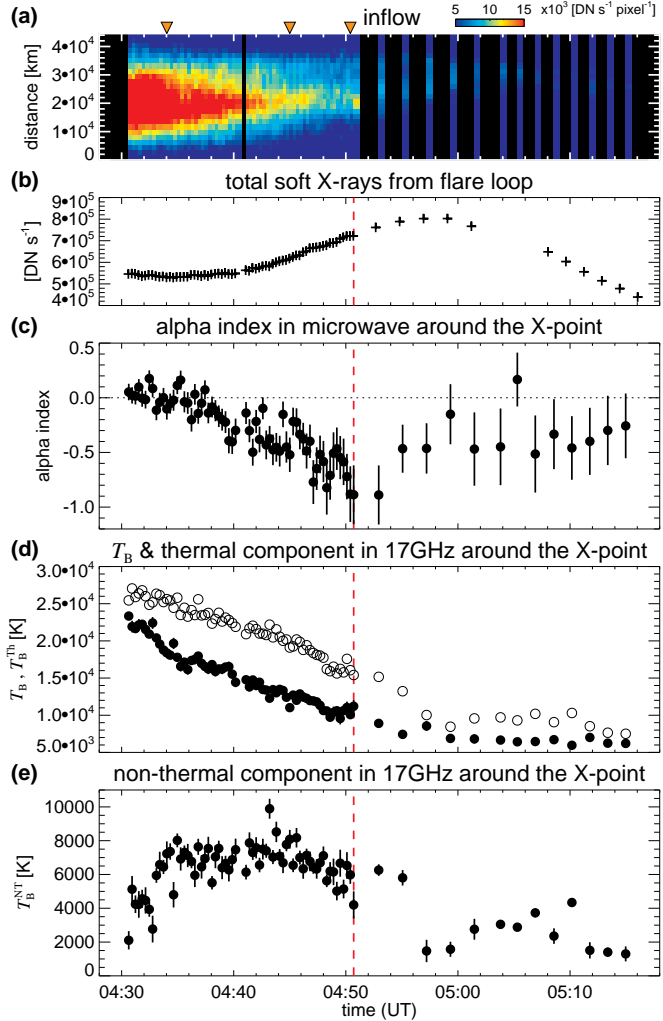}
\caption{
Temporal evolution of the reconnection inflow and parameters around the X-point
derived from the microwave observations.
{\bf (a)} Time-distance plot generated by stacking soft X-ray intensities (with the AlMg filter)
along the yellow lines in Figure~\ref{fig:data}~(b)--(d) in order of time.
The end point ``A" of the yellow line corresponds to the origin in distance (0~km) in this plot.
Black portions correspond to data gaps in the soft X-ray observations.
Orange triangles in the top part of this panel indicate the times when the images in Figure~\ref{fig:data} were taken.
{\bf (b)} Total soft X-ray intensity from the bright closed flare loop taken with the Be119 filter.
{\bf (c)} Temporal evolution of the alpha index around the X-point with error bars (1$\sigma$).
The main cause of this error is the uncertainty in removing the aliasing pattern in the 34~GHz maps
(see Appendix~\ref{sec:Calibration of 34 GHz data} for details).
{\bf (d)} Observed brightness temperature at 17~GHz ($T_\mathrm{B}$; white circles)
and expected brightness temperature from the thermal component ($T_\mathrm{B}^\mathrm{Th}$; black circles)
deduced from the temperature and column emission measure (Figure~\ref{fig:sxr}).
1-$\sigma$ errors in $T_\mathrm{B}^\mathrm{Th}$ from the errors shown in Figure~\ref{fig:sxr} are
also displayed (but there are mostly within the black circles).
{\bf (e)} Estimated 17~GHz brightness temperature of non-thermal emission originating around the X-point
together with 1-$\sigma$ error bars from subtraction of $T_\mathrm{B}^\mathrm{Th}$ from $T_\mathrm{B}$.
The vertical red dashed lines indicate the time of the plasmoid ejection.
\label{fig:microwave}}
\end{figure}

\begin{figure}
\epsscale{.55}
\plotone{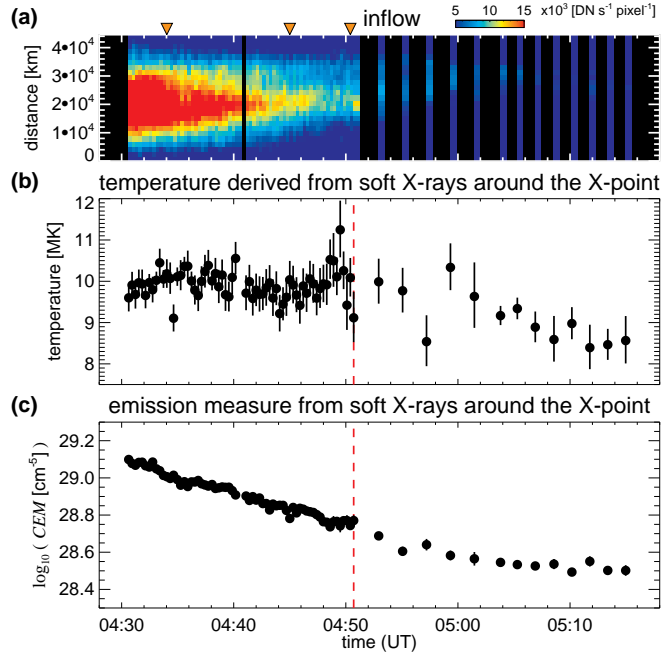}
\caption{
Reconnection inflow and temporal evolution of thermal plasma around the X-point derived from the soft X-ray observations.
{\bf (a)} The same time-distance plot as in Figure~\ref{fig:microwave}~(a), showing the inflow pattern.
{\bf (b) and (c)} Filter-ratio temperature and column emission measure around the X-point with 1-$\sigma$ error bars
(dominated by soft X-ray photon noise) obtained from a pair of AlMg and Be119 filters.
The vertical red dashed lines indicate the time of the plasmoid ejection.
\label{fig:sxr}}
\end{figure}

\begin{figure}
\epsscale{.50}
\plotone{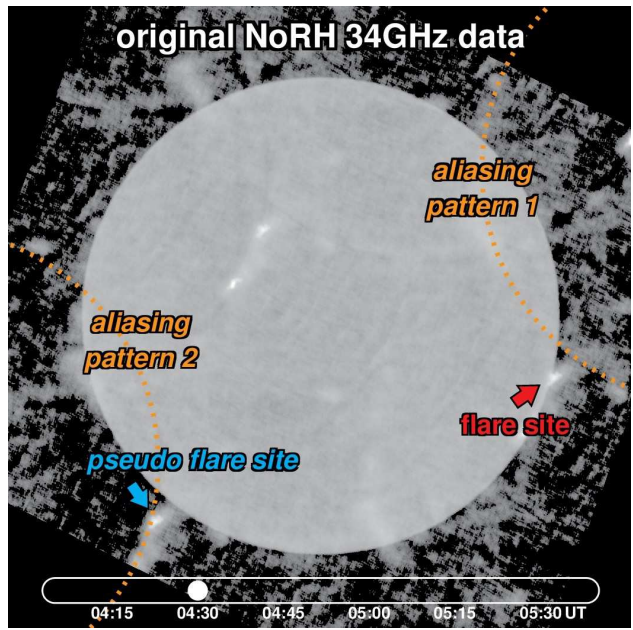}
\caption{
Example of original 34~GHz map from NoRH (a snap shot of Movie~5 at 04:30:00~UT).
The analyzed flare site is shown with a red arrow. The solar limbs of
aliasing patterns~1 and 2 are indicated with orange dotted circular arcs.
The pseudo flare site produced by the aliasing effect is shown with a blue arrow.
\label{fig:34 G data}}
\end{figure}

\begin{figure}
\epsscale{.50}
\plotone{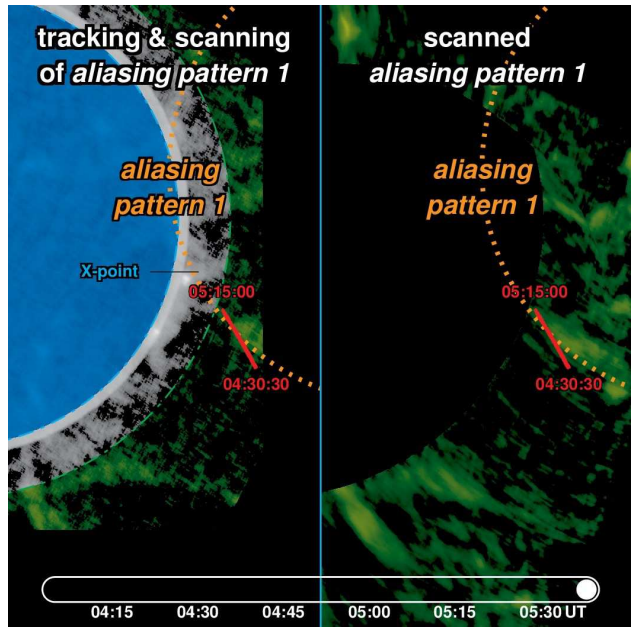}
\caption{
Aliasing pattern in the 34~GHz maps (a snap shot of Movie~6 at 05:38:00~UT).
The green-colored zone in the left side of the figure indicates the region used for
extracting AP1 as described in the text. The right side shows the extracted AP1 in green.
The red line on both sides indicates the trajectory along which the corresponding points
on AP1 pass through the X-point. In the figure, the on-disk portion used for extracting AP1
before it passed the flare site is shown in blue (see text for details). 
\label{fig:track and scan}}
\end{figure}

\begin{figure}
\epsscale{.60}
\plotone{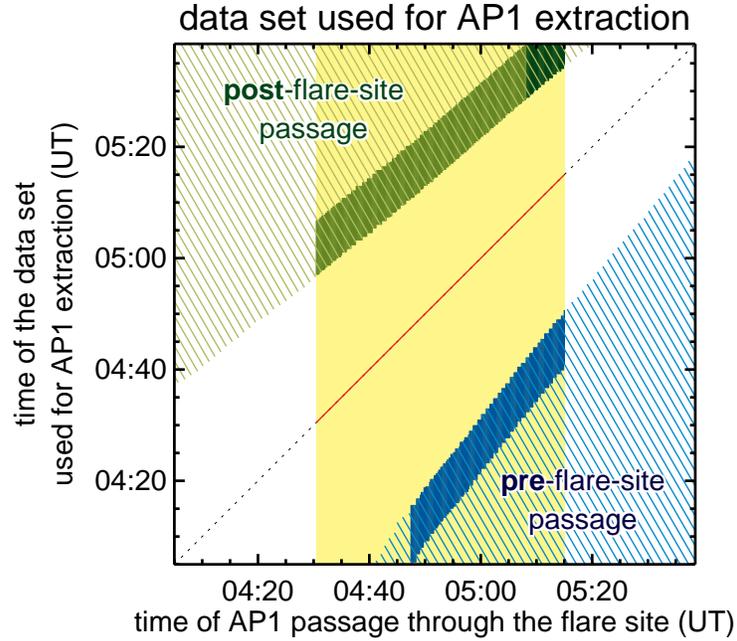}
\caption{
Data set used for AP1 extraction.
The horizontal and vertical axes show the time of AP1 passage through the flare site and
the time of the 34~GHz map used for extracting AP1.
The yellow area indicates the time period when the flare was analyzed in the paper.
The red line denotes the ideal case where AP1 passing through the flare site were to be extracted
without any time gap, which was not possible in our analysis. The green- and blue-hatched areas show
the time when the AP1 extraction was possible using the green- and blue-colored areas
in the left-hand side of Figure~\ref{fig:track and scan}, respectively.
The region in light green color indicates that the data sets used for extracting AP1 included 60 maps
of post-flare-site passage. The small dark green area indicates that less than 60 maps were used.
The blue colored region shows the data set that used 60 maps of pre-flare-site passage.
\label{fig:cal dataset}}
\end{figure}

\begin{figure}
\epsscale{.50}
\plotone{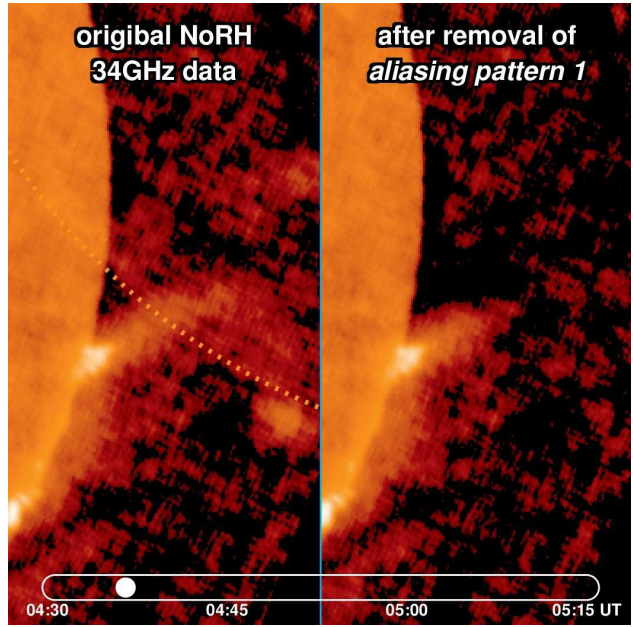}
\caption{
Pre- and post-removal of AP1 around the flare site at 34~GHz (a snap shot of Movie~7 at 04:36:32~UT).
The left-hand panel presents an original 34~GHz map while the right-hand panel the same map after removal of AP1.
\label{fig:cal result}}
\end{figure}

\begin{figure}
\epsscale{.70}
\plotone{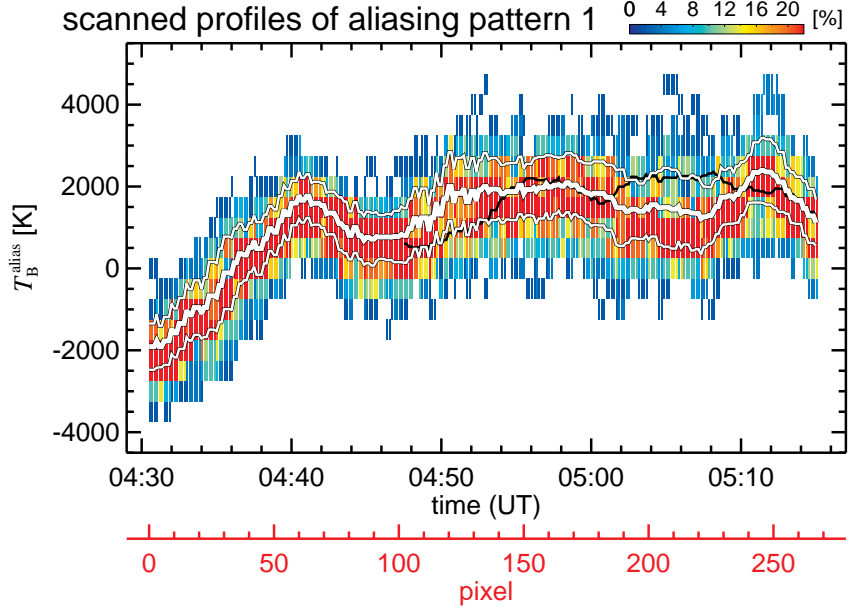}
\caption{
Brightness temperature profile of AP1 for points that pass through the X-point region.
The profile is obtained along the red line shown in Figure~\ref{fig:track and scan},
by determining AP1 using the green-colored zone in Figure~\ref{fig:track and scan}
after it passed the X-point region of the flare.
The red and black horizontal axes show the locations along the red line and
the time when the corresponding point in AP1 passed through the X-point, respectively.
The fractional distribution of the brightness temperature for each horizontal axis location is shown in color.
The thick and thin white lines give the average brightness for each horizontal axis location and
the associated range of standard deviation, respectively.
The black line in the panel (data were only available after 04:47:30~UT) indicates
the brightness temperature profile of AP1 determined from the on-disk portion of the microwave Sun,
before it passed the X-point region.
\label{fig:cal check}}
\end{figure}

{\bf Movie 3:}
Evolution of the flare in soft X-rays and in microwave frequencies.
Top left panel: the same time-distance plot as in Figure~\ref{fig:microwave}~(a) and Figure~\ref{fig:sxr}~(a),
with the corresponding time for each video frame indicated by a vertical yellow line.
Top middle panel: soft X-ray images taken with the AlMg filter of {\it Yohkoh}/SXT.
Top right panel: plasma temperature derived by the filter-ratio method from pairs of AlMg and Be119 filter images from {\it Yohkoh}/SXT.
Bottom panels (from left to right): EUV full-sun images, brightness temperature maps at 17~GHz, brightness temperature maps at 34~GHz,
and the spatial distribution of the alpha index derived from 17~GHz and 34~GHz flux densities.
The EUV images were observed with {\it SOHO}/EIT in 195 {\AA}.
The white square in the EUV full-sun image indicates the field-of-view of the other panels.
The microwave data were taken with the Nobeyama Radioheliograph.

{\bf Movie 4:}
Evolution of the flare observed in soft X-rays with the AlMg filter of {\it Yohkoh}/SXT (orange)
overlaid with regions of negative alpha index from 17~GHz and 34~GHz microwave flux densities
from the Nobeyama Radioheliograph (green; from 04:30:38 UT to 05:14:58 UT).
Negative alpha indices of less than $-0.1$ with a 1-$\sigma$ error less than 0.5 are indicated in green,
which saturates at an alpha index of $-0.5$ ($-0.1$: transparent green, $-0.5$: opaque green).

{\bf Movie 5:}
Original brightness temperature maps at 34~GHz from NoRH that contain aliasing patterns.
The flare site analyzed in this study is shown with a red arrow.
The solar limbs of aliasing patterns 1 and 2 are indicated with orange dotted circles.
The pseudo flare site produced by the aliasing effect is shown with a blue arrow.

{\bf Movie 6:}
Aliasing pattern in the 34 GHz maps.
The green-colored zone in the left side of the video shows the region used for extracting AP1.
The right side shows the extracted AP1.
The red line on both sides indicates the trajectory
along which the corresponding points on AP1 pass through the X-point.

{\bf Movie 7:}
Pre- and post-removal of AP1 around the flare site at 34~GHz.
The left-hand panel presents original 34~GHz maps while the right-hand panel the same maps after removal of AP1.



\clearpage

\begin{table}
\begin{center}
\caption{
The power-law index of the microwave flux density spectra (alpha index).
\label{tbl:alpha index}}
\begin{tabular}{ccc}
\tableline\tableline
Emission mechanism & \multicolumn{2}{c}{alpha index} \\
\cline{2-3}
 & Optically thick & Optically thin \\
\tableline\tableline
Non-thermal gyro-syncrotoron          & $+2.9 \pm 0.1$ & $-1.5$ for $\delta_{\mu}=3$\tablenotemark{a} \\
from mildly-relativistic electrons    &                & $-4.2$ for $\delta_{\mu}=6$ \\
\tableline
Thermal gyro-syncrotoron (Gyro-resonance) & $+2$           & $-8$ \\
\tableline
Thermal Bremsstrahlung   & $+2$           & $0$ \\
\tableline
\end{tabular}
\tablenotetext{a}{$\delta_{\mu}$ is the electron spectral index.}
\tablecomments{This table is the summary of Figure 2 in \citet{dul85}.}
\end{center}
\end{table}


\clearpage

\clearpage

\clearpage

\clearpage

\clearpage

\clearpage

\clearpage

\clearpage


\begin{thebibliography}{}
\bibitem[Asai et al.(2013)]{asa13} Asai, A. et al.  2013, \apj, 763, 87
\bibitem[Aschwanden et al.(1996)]{asc96} Aschwanden, M. J., Kosugi, T., Hudson, H. S., Wills, M. J. \& Schwartz, R. A.  1996, \apj, 470, 1198--1217
\bibitem[Bastian(2002)]{bas02} Bastian, T. S.  2002, Astron. Nachr. 323, 271--276
\bibitem[Drake et al.(2006)]{dra06} Drake, J. F., Swisdak, M., Che, H. \& Shay, M. A.  2006, \nat, 443, 553--556
\bibitem[Dulk(1985)]{dul85} Dulk, G. A.  1985, Ann. Rev. Astron. Astrophys., 23, 169--224
\bibitem[Fletcher \& Hudson(2008)]{fle08} Fletcher, L. \& Hudson, H. S.  2008, \apj, 675, 1645--1655
\bibitem[Hara et al.(2011)]{har11} Hara, H. et al.  2011, \apj, 741, 107
\bibitem[Imada et al.(2013)]{ima13} Imada, S. et al.  2013, \apjl, 776, L11
\bibitem[Koyama et al.(1995)]{koy95} Koyama, K. et al.  1995, \nat, 378, 255--258
\bibitem[Krucker et al.(2010)]{kru10} Krucker, S. et al.  2010, \apj, 714, 1108--1119
\bibitem[Krucker et al.(2013)]{kru13} Krucker, S. et al.  2013, Proceedings of the SPIE, 8862, 88620R
\bibitem[Lin \& Hudson(1976)]{lin76} Lin, R. P. \& Hudson, H. S.  1976, \solphys, 50, 153--178
\bibitem[Lin et al.(1996)]{lin96} Lin, R. P. et al.  1996, \grl, 23, 1211--1214
\bibitem[Litvinenko(1996)]{lit96} Litvinenko, Y. E.  1996, \apj, 462, 997--1004
\bibitem[Masuda et al.(1994)]{mas94} Masuda, S., Kosugi, T., Hara, H., Tsuneta, S. \& Ogawara, Y.  1994, \nat, 371, 495--497
\bibitem[Mushotzky(1977)]{mus77} Mushotzky, R. F.  1977, \nat, 265, 225--226
\bibitem[Nakajima et al.(1994)]{nak94} Nakajima, H. et al.  1994, Proc. IEEE. 82, 705--713
\bibitem[Narukage \& Shibata(2006)]{nar06} Narukage, N. \& Shibata, K.  2006, \apj, 637, 1122--1134
\bibitem[{\O}ieroset et al.(2002)]{oie02} {\O}ieroset, M., Lin, R. P., Phan, T. D., Larson, D. E. \& Bale, S. D.  2002, \prl, 89, 195001-1-4
\bibitem[Oka et al.(2010)]{oka10} Oka, M. et al.  2010, \apj, 714, 915
\bibitem[Ohyama \& Shibata(1998)]{ohy98} Ohyama, M. \& Shibata, K.  1998, \apj, 499, 934--944
\bibitem[Priest(1982)]{pri82} Priest, E. R.  1982, Solar Magnetohydrodynamics, vol. 21 of Geophysics and Astrophysics Monographs, (Dordrecht: Reidel, Boston)
\bibitem[Pritchett(2006)]{pri06} Pritchett, P. L.  2006, \grl, 33, L13104
\bibitem[Sakao et al.(1996)]{sak96} Sakao, T. et al.  1996, Advances in Space Research, 17, 67--70
\bibitem[Sakao et al.(1998)]{sak98} Sakao, T., Kosugi, T. \& Masuda, S.  1998, Astrophysics and space science library, 229, 273--284
\bibitem[Sakao et al.(2012)]{sak12} Sakao, T. et al.  2012, Proceedings of the SPIE, 8443, 84430A
\bibitem[Selhorst et al.(2005)]{sel05} Selhorst, C. L., Silva, A. V. R., \& Costa, J. E. R.  2005, \aap, 433, 365--374
\bibitem[Shibata \& Tanuma(2001)]{shi01} Shibata, K. \& Tanuma, S.  2001, Earth Planets Space, 53, 473--482
\bibitem[Shibata \& Magara(2011)]{shi11} Shibata, K. \& Magara, T.  2011, Living Rev. Solar Phys., 8, 6
\bibitem[Somov \& Kosugi(1997)]{som97} Somov, B. V. \& Kosugi, T.  1997, \apj, 485, 859--868
\bibitem[Su et al.(2013)]{su13} Su, Y. et al.  2013, Nature Physics, 9, 489--493
\bibitem[Takano et al.(1997)]{tak97} Takano, T. et al.  1997, Coronal Physics from Radio and Space Observations; Proceedings of the CESRA Workshop held in Nouan le Fuzelier, 483, 183--191
\bibitem[Takeda(2011)]{tak11} Takeda, A.  2011, \solphys, 273, 295--306
\bibitem[Tsuneta et al.(1991)]{tsu91} Tsuneta, S. et al.  1991, \solphys, 136, 37--67
\bibitem[Tsuneta et al.(1992)]{tsu92} Tsuneta, S. et al.  1992, \pasj, 44, L63--L69
\bibitem[Tsuneta \& Naito(1998)]{tsu98} Tsuneta, S. \& Naito, T.  1998, \apjl, 495, L67--L70
\bibitem[Yokoyama et al.(2001)]{yok01} Yokoyama, T., Akita, K., Morimoto, T., Inoue K., \& Newmark, J.  2001, \apjl, 546, L69--L72
\bibitem[Yokoyama et al.(2002)]{yok02} Yokoyama, T., Nakajima, H., Shibasaki, K., Melnikov, V. F. \& Stepanov, A. V.  2002, \apjl, 576, L87--L90
\end{thebibliography}
\end{document}